\documentclass[amsmath,amssymb,aps,showpacs]{revtex4}
\usepackage{graphicx,subfigure}
\usepackage{array}
\usepackage{mathtools}
\newcommand{\bigO}[1]{O(#1)}
\def\ket#1{\mathinner{|{#1}\rangle}}
\def\bra#1{\mathinner{\langle{#1}|}}
\newcommand{\defas}{\coloneqq}

\newcommand{\ketbra}[3][]{\mathinner{|#2\rangle\langle #3|}_{#1}}
\newcommand{\proj}[2][]{\ketbra[#1]{#2}{#2}}
\newcommand{\abs}[1]{\left|#1\right|}
\renewcommand{\i}{\mathrm{i}}
\newcommand{\e}{\mathrm{e}}
\newcommand{\braket}[2]{\langle #1 | #2 \rangle}

\newcommand{\bof}{\tan^2 \theta}
\newcommand{\lep}{P_{K-1}\left(\frac{1+\tan^2 \theta}{1-\tan^2 \theta}\right)}
\newcommand{\lem}{P_{K-1}\left(\frac{1-\tan^2 \theta}{1+\tan^2 \theta}\right)}

\begin{document}
\title{Decoherent histories of quantum searching}

\author{Wim van Dam}
\email[Email: ]{vandam@cs.ucsb.edu}
\affiliation{%
Department of Computer Science, Department of Physics, University of California, Santa Barbara, CA 93106, USA
}

\author{Hieu D.\ Nguyen}
\email[The authors are listed in alphabetical order; HDN is the primary and corresponding author. Email: ]{hdn@physics.ucsb.edu}
\affiliation{%
Department of Physics, University of California, Santa Barbara, CA 93106, USA
}%

\date{\today}

\begin{abstract}
The theory of decoherent histories is an attempt to derive classical physics from positing only quantum laws at the fundamental level without notions of a classical apparatus or collapse of the wave-function. Searching for a marked target in a list of $N$ items requires $\Omega(N)$ oracle queries when using a classical computer, while a quantum computer can accomplish the same task in $\bigO{\sqrt{N}}$ queries using Grover's quantum algorithm. We study a closed quantum system executing Grover algorithm in the framework of decoherent histories and find it to be an exactly solvable model, thus yielding an alternate derivation of Grover's famous result. We also subject the Grover-executing computer to a generic external influence without needing to know the specifics of the Hamiltonian insofar as the histories decohere. Depending on the amount of decoherence, which is captured in our model by a single parameter related to the amount of information obtained by the environment, the search time can range from quantum to classical. Thus, we  identify a key effect induced by the environment that can adversely affect a quantum computer's performance and demonstrate exactly how classical computing can emerge from quantum laws.
\end{abstract}

\maketitle

\section{Introduction.}
Quantum computer science has come to know numerous examples in which a quantum computer can outperform its classical counterpart \cite{chuang}. Among the celebrated algorithms is that a quantum computer can search for a marked element from a list of $N$ items using the oracle only $\bigO{\sqrt{N}}$ number of times rather than the classical $\Omega(N)$ \cite{grover}. The quantum algorithm, which gives the picture of a state initially being an equal superposition of all candidates rotating onto the target, is entirely dissimilar to the simple process of elimination of classical physics. Indeed even a classical computer operates according to quantum laws, yet we do not understand what happens differently that gives a quantum computer its speed-up. Numerous elements including entanglement, interference, and quantum correlation have been suspected as the key resource responsible for the quantum enhancement \cite{josza, bruss,lloyd,meyer, cast}. On a related note cosmologists have invented the theory of decoherent histories as a way to derive classical physics from fundamental quantum laws\cite{hartle, grif2}. In this paradigm there is no collapse of the wave-function or classical measuring apparatus. The Schrodinger equation is the only dynamical law, yet deterministic and classical laws governing coarse-grained observables can arise. The theory of decoherent histories has been applied to quantum information processing to identify information that can be made classical without losing substantially quantum computing power \cite{poulin}. Unfortunately only a few dynamical models have been solved exactly to demonstrate classical emergence \cite{brun, brun2, hartle-path}. 

In the current paper we apply the theory of decoherent histories to Grover's algorithm as a particular example illustrating how classical computational power can arise within a closed quantum system. We are able to compute the main figure of merit in closed form and recover the well known quadratic speed-up. Furthermore, changing the Grover dynamics to a generic unitary about which we need not know anything except that it makes records about how the target is found available, we recover classical search time. Although we make a number of simplifying assumptions in the derivation, our results show how computing power depends on the decoherence of histories. In the next section we will set up the search problem within the theory of decoherent histories and derive the famous quadratic speed-up. Then we present the modified Grover dynamics under a generic environmental influence and derive the search time as a function of decoherence. 
\section{Quantum search.}
Consider the $N$ dimensional Hilbert space of $N$ independent spins spanned by $\{\ket{m}\}_{m=1}^N$ where $\ket{m}$ denotes spin $m$ being up while the rest of them down. The marked element $w$ which is to be found is distinguished in the following manner in the Hamiltonian
\begin{equation}
H_w = -\sum_{m=1}^N (-1)^{\delta_{mw}} \sigma_m -(N-2)
\end{equation}
where $\sigma_m$, in the ordered basis $\{\uparrow, \downarrow\}$, is the tensor product of $\begin{pmatrix}
1 & 0\\
0 &-1
\end{pmatrix}$
on the $m$-th spin and identity on the rest. This Hamiltonian $H_w$ has the desired spectrum of all the ground states having energy $0$ except for the marked state having energy $2$.  To this Hamiltonian we add the driving Hamiltonian $2 \proj{s}$, where 
\begin{equation}
\ket{s} = \frac{1}{\sqrt{N}}\sum_{m=1}^N \ket{m}
\end{equation}
and consider henceforth the total Hamiltonian
\begin{equation}
H = H_w + 2 \proj{s}. 
\end{equation}
The unitary evolution driven by $H$ will rotate the initial state $\ket{s}$ onto the target $\ket{w}$ in a time proportional to $O(\sqrt{N})$ \cite{farhi}. We will make $t$ short enough so that $U(t)$ is equivalent to $1$ application of the oracle in discrete time and apply the oracle for a total of $K$ times. With the unity partition of projectors
\begin{align}
P_1 &= \proj{w} \\
P_0 &= 1 - P_1,
\end{align}
generally
\begin{align}
U(tK)\ket{s} & = \label{eq:Utk}
((P_1+P_0)U(t))^{K}\ket{s} \\
& = \sum_{\alpha \in \{0,1\}^K} C_\alpha \ket{s}. \label{eq:sumCa}
\end{align}
where
\begin{equation}\label{def:Ca}
C_{\alpha} \defas P_{\alpha_K} U(t) \cdots P_{\alpha_i} U(t) \cdots P_{\alpha_1} U(t).
\end{equation}
 While the well-known perspective shows Grover search to be a rotation, the theory of decoherent histories considers the unitary evolution as a sum of history branches $C_\alpha \ket{s}$ as in Equation~\eqref{eq:sumCa}. Each query is an attempt to ask the question ``Is the target found,'' and a history is a string of answers ``yes'' or ``no'' to the question, as represented by the projectors $P_1$ and $P_0$, respectively. For brevity, when successive components of $\alpha$ are alike, we can collect them into a streak. For example, if a history consists of $0$'s for the first $s_1$ events, $1$'s for the next $s_2$ events, and so forth up to $1$'s for the last $s_n$ events, then we would write it as $\alpha = 1^{s_n}\dots 1^{s_2}0^{s_1}$, where $n$ is the number of streaks in the history with $1 \leq n \leq K$ and $s_j$ the length of streak $j$. The projectors at each time slice provide for an algebra of events, and each $\alpha$ has the same meaning as a path in a random walk, yet we may not be able to assign probabilities to them. Let us define the following quantities in terms
of $N$ and $w$:
\begin{align}
\ket{\xi} &\defas \frac{1}{\sqrt{N-1}}\sum_{m = \{1,\dots, N\} \setminus \{w\}} \ket{m},\\
x &\defas \frac{1}{\sqrt{N}},\\
%-\i \sqrt{\frac{N-1}{N}}\sin xt.
f &\defas \cos xt - \i x \sin xt = \abs{f}{\e^{\i \phi}}\\
b &\defas -i \sqrt{\frac{N-1}{N}} \sin xt
\end{align}
Starting from the definition of operators $C_\alpha$ of Equation~\eqref{def:Ca}, one can show that
\begin{widetext}
\begin{equation}\label{main}
C_\alpha \ket{s} =  \abs{f}^{K-n}b^{n-1}
\begin{cases}
\displaystyle\exp\Bigl( \i \phi \Bigl( \sum_{k=\textrm{even}}\alpha_k - \sum_{k=\textrm{odd}}\alpha_k\Bigr) \Bigr) 
\Bigl(f^*\sqrt{\frac{N-1}{N}}+bx\Bigr)\ket{w}, &\mbox{if } \alpha = 1^{s_n},\dots,0^{s_1}, n = \textrm{even}\\
\displaystyle
\exp\Bigl( \i \phi \Bigl( \sum_{k=\textrm{odd}}\alpha_k - \sum_{k=\textrm{even}}\alpha_k\Bigr) \Bigr)\Bigl(b\e^{-\i \phi}\sqrt{\frac{N-1}{N}} + \abs{f} x\Bigr)\ket{w}, &\mbox{if } \alpha =1^{s_n},\dots, 1^{s_1}, n = \textrm{odd}\\
\displaystyle
\exp\Bigl( \i \phi \Bigl( \sum_{k=\textrm{odd}}\alpha_k - \sum_{k=\textrm{even}}\alpha_k\Bigr)\Bigr)
\Bigl(b\sqrt{\frac{N-1}{N}} + fx\Bigr)\ket{\xi}, &\mbox{if } \alpha =0^{s_n},\dots, 1^{s_1}, n = \textrm{even}\\
\displaystyle
\exp\Bigl(\i \phi \Bigl( \sum_{k=\textrm{even}}\alpha_k - \sum_{k=\textrm{odd}}\alpha_k\Bigr)\Bigr)
\Bigl(\abs{f}\sqrt{\frac{N-1}{N}} + \e^{\i \phi} bx\Bigr)\ket{\xi}, &\mbox{if } \alpha =0^{s_n},\dots, 0^{s_1}, n = \textrm{odd}
\end{cases}
\end{equation}
\end{widetext}
Clearly there exist $\alpha \neq \alpha'$ such that the decoherence functional
\begin{equation}
D(\alpha, \alpha') \defas \bra{s}C^{\dagger}_{\alpha'}C_{\alpha}\ket{s} \label{def:med}
\end{equation}
does not vanish. If the decoherence functional vanishes $\forall \alpha \neq \alpha'$, we would have medium decoherence, which is equivalent to existence of records  \cite{grif,halliwell}.
%\emph{cite hartle}.\emph{cite dowker}
Moreover, probabilities would be assigned to each $\alpha$, thus making the Grover search a classical stochastic process.
\subsection{Time of Grover's Search Algorithm}
We will now derive the time required to find the target. In the regime $N \gg 1$ we have $xt \in \bigO{x} = \bigO{1/\sqrt{N}} = o(1), \abs{b} \approx xt, \abs{f} \approx 1$, and $\abs{\phi} \approx 1/N$. Moreover, the total time for which we study the process is proportional to $K$, so long as $K \lesssim \sqrt{N}$, the phases are of order $\phi K \approx 1/\sqrt{N}$ and therefore negligible. As a result, Equation~\eqref{main} is simplified to
\begin{equation}\label{c_approx}
C_\alpha \ket{s} \approx \abs{f}^{K-n}b^{n-1}  
\begin{cases}
\displaystyle
S(n)\ket{w}, & \mbox{if }\alpha = 1^{s_n}\ldots \\
\displaystyle
F(n) \ket{\xi}, & \mbox{if }\alpha =0^{s_n}\ldots
\end{cases}
\end{equation}
where 
\begin{equation}\label{eq:defS}
S(n) = \begin{cases}
f^*, &n = \textrm{even}\\
b + \abs{f} x, &n = \textrm{odd},
\end{cases} 
\end{equation}
and
\begin{equation}\label{eq:defF}
F(n) = \begin{cases}
b+fx, &n=\textrm{even}\\ 
\abs{f}+ bx, &n=\textrm{odd}.
\end{cases}
\end{equation}
From Equations~\eqref{eq:sumCa} and \eqref{c_approx}, the probability of success after time $tK$ is the amplitude square of the following branch
\begin{align}
& P_1 U(tK) \ket{s} = \sum_{\alpha \in \{1\}\times \{0,1\}^{K-1}} C_\alpha \ket{s}\\
&\approx \sum_{n=1}^{K} \binom{K -1}{n-1}\abs{f}^{K-n}b^{n-1}S(n)\ket{w},
\end{align}
where $\binom{K -1}{n-1}$ is the number of histories with $n$ streaks for a given $K$. Moreover, since $\abs{f}^2 + \abs{b}^2 = 1$, there exists a $\theta$ such that
\begin{align}
\sin \theta &= \abs{b} \, \textrm{and}\\
\cos \theta &= \abs{f}
\end{align}
Using the identity found in \cite{table} to compute the sum above, we find
\begin{equation}\label{suc}
P_1 U(tK) \ket{s} \approx -\i S(\textrm{even})\sin (K-1)\theta + S(\textrm{odd})\cos (K-1)\theta
\end{equation} 
Similarly the sum of the branches that ends in failure is
\begin{equation}\label{fail}
P_0 U(tK) \ket{s} \approx -\i F(\textrm{even})\sin (K-1)\theta + F(\textrm{odd})\cos (K-1)\theta
%\sum_{n=1}^K \binom{K -1}{n-1}\abs{f}^{K-n}b^{n-1} F(n)\ket{\xi} 
\end{equation}
Note that $\tan \theta = \bigO{1/\sqrt{N}}$ and is small. Thus the number of oracle queries sufficient to find the target is when $(K-1) \theta = \pi/2$, yielding $K = \bigO{1/x} = \bigO{\sqrt{N}}$ consistent with Grover's well-known result. 
\section{Classical search.}
We now expand the treatment so as to include classical as well as quantum results in one framework. Consider the tensor product space of the system of $N$ spins that executes Grover algorithm and its environment. Let the initial state be $\ket{s} \otimes \ket{\Psi_e}$, where $\ket{\Psi_e}$ is the state of the environment, and $\tilde{U}(t)$ be the unitary evolution on the composite. Between applications of the oracle, we still ask the question ``Is the target found,'' and the answers ``yes'' and ``no'' are represented respectively as
\begin{align}
\Pi_1 &= \proj{w} \otimes 1\\
\Pi_0 &= (1-\proj{w}) \otimes 1
\end{align}
Again denoting histories by $\alpha \in \{0,1\}^K$, we define the history branch operators similarly to Equation~\eqref{def:Ca} as
\begin{equation}\label{def:Ga}
G_{\alpha} \defas \Pi_{\alpha_K} \tilde{U}(t) \cdots \Pi_{\alpha_i} \tilde{U}(t) \cdots \Pi_{\alpha_1} \tilde{U}(t).
\end{equation}
Rather than specifying the interaction Hamiltonian, we make the following assumptions:
\begin{enumerate}
\item $G_\alpha \ket{s}\otimes \ket{\Psi_e} = \frac{1}{A} C_\alpha \ket{s} \otimes \ket{e_\alpha},$ where the environment states $\ket{e_\alpha}$ are assumed to be normalized, and $A$ is a normalizing factor independent of $\alpha$ such that
\begin{equation}
\abs{\sum_\alpha G_\alpha \ket{s}\otimes \ket{\Psi_e}}^2 = 1. 
\end{equation}
\item \label{eq:va}
$\braket{e_{\alpha'}}{e_\alpha} = 0$
for histories $\alpha$ and $\alpha'$ having different numbers of streaks.
\item \label{eq:defn-d}
\begin{equation}\delta \defas \frac{1}{\binom{K-1}{n-1}(\binom{K-1}{n-1}-1)}\sum_{\substack{\alpha \neq \alpha'\\ \textrm{histories of }n\textrm{ streaks}}}\braket{e_{\alpha'}}{e_\alpha}
\end{equation} to be independent of $n$.
\end{enumerate}
Up to this point we have left the interaction with the environment completely unspecified, and it could be so destructive that the system of $N$ spins no longer executes the search. Therefore, the first assumption is necessary since it restricts the interaction in a way that Grover search is still being executed with the additional complication that each history branch is now correlated with a state in the environment. Records of which branch are kept by the environment in the states $e_{\alpha}$, but they may not be orthogonal to each other. Therefore, the information of which branch is not necessarily available, but the second assumption makes available the information about a branch's number of streaks. Since $\delta$ is defined to be proportional to the average cosine of the angle between two branches having a given number of streaks, if the branches are orthogonal or have randomized phases relative to each other, then $\delta = 0$. Either case is a signature of classical behavior, and we take $\delta$ to be a measure of decoherence with $0$ corresponding to classical and $1$ to possibly quantum. Finally, assuming $\delta$ to be independent of $n$ is for convenience yet is general enough to reveal the quantum-ness responsible for the square root speed-up. Using the last two assumptions and the series representation of Legendre polynomials found in \cite{nist}, we obtain
\begin{equation}
% \begin{split}
% &
\abs{A}^2 \approx \abs{f}^{2K} \left[ \delta \left(1- \tan^2\theta\right)^{K-1}\lep % \right. \\
% &\left. 
+ (1-\delta) \left(1 + \tan^2 \theta\right)^{K-1}\right],
% \end{split}
\end{equation}
where $P_{K-1}$ is a Legendre polynomial. After $K$ iterations of the oracle, similar to the Grover analysis, the probability of success is
\begin{widetext}
\begin{align}
\mathrm{Pr}(\textrm{Success}) &= \abs{\sum_{\alpha \in \{1\}\times\{0,1\}^{K-1} } G_\alpha\ket{s}\otimes \ket{\Psi_e}}^2 \\
&\approx \frac{\abs{f}^{2K}}{\abs{A}^2}(1-\delta) \left[ \frac{(1+\bof)^K - (1-\bof)^K}{2} \right] + \frac{\abs{f}^{2K}}{\abs{A}^2} \delta\times \nonumber \\
&\left[ (1-\bof)^{K-1} \left(\frac{1+\bof}{2}\right) \lep -(1+\bof)^{K-1}\left(\frac{1-\bof}{2}\right)\lem \right]  \label{eq:qc}
\end{align}
\end{widetext}
When $\delta = 0$, the above equation simplifies
\begin{equation}
\mathrm{Pr}(\textrm{Success}) = \frac{1+\bof}{2}\left(1- \left(\frac{1-\bof}{1+\bof}\right)^K \right)
\end{equation}
Hence, the probability of finding the target is of order unity when $K = \bigO{N}$, which is the classical search time. When $\delta = 1$, we have
\begin{equation}
% \begin{split} &
\mathrm{Pr}(\textrm{Success}) = \frac{1+\bof}{2}- \frac{1-\bof}{2}
% &
\left( \frac{1+\bof}{1-\bof}\right)^{K-1}\frac{\lem}{\lep}.
% \end{split}
\end{equation}
Since the expansion of the Legendre polynomial is $\lem \approx 1 - K(K-1)\bof$, the probability of success is of order unity when $K = \bigO{\sqrt{N}}$, which is quantum time. As shown in Equation~\eqref{eq:qc} the probability of success in general is a convex combination between classical and quantum search times. 
\section{Conclusion.}
We find studying the quantum search in the framework of decoherent histories to be beneficial in many ways. Firstly, the Grover system makes an instructive toy application for the theory of decoherent histories, which allows for an alternative derivation of the quantum search time. Secondly, we see an exactly solvable demonstration of how classical computing can arise out of quantum systems. After each application of the oracle, we ask the question ``Is the target found'', and a search history, which is a series of answers ``yes'' or ``no'' to the question, is a vector, whose orientation relative to each other determines the search time and is affected by the environment. In the pure Grover dynamics case, all the history branches that end in success are phase coherent and aligned in one direction as evident in Equation~\eqref{main}. Thus, no information about the search is made available. The role of the environmental interaction is that it can disrupt the history branches' orientations to affect the search time. In contrast to the pure Grover dynamics, our model shows that a little bit of information, namely the number of streaks in the search history, can be leaked out to the environment without sacrificing the quadratic speed-up. However, if the history branches have randomized phases or are orthogonal to each other, which is equivalent to the environment keeping complete records of the search, then the search time will be classical scaling. For a general level of decoherence the search time is a convex combination between classical and quantum as given in Equation~\eqref{eq:qc}. Thus, applying the theory of decoherent histories to the search problem allows us to see how the same quantum laws at the fundamental level can give rise to both quantum and classical computing powers depending on the amount of information obtained by the environment. 

\paragraph*{Acknowledgements:}
This material is based upon work supported by the National Science Foundation under
Grant No.~0747526.

\end{document}